\documentclass[letterpaper]{article} 
\usepackage{aaai24}  
\usepackage{times}  
\usepackage{helvet}  
\usepackage{courier}  
\usepackage[hyphens]{url}  
\usepackage{graphicx} 
\usepackage{natbib}  
\usepackage{caption} 
\usepackage{algorithm}
\usepackage{algorithmic}
\usepackage{amssymb}
\usepackage{amsmath}
\usepackage{nameref}

\usepackage{booktabs, multirow} 
\usepackage{soul}
\usepackage[table]{xcolor} 
\usepackage{changepage,threeparttable} 

\usepackage{newfloat}
\usepackage{listings}
\DeclareCaptionStyle{ruled}{labelfont=normalfont,labelsep=colon,strut=off} 
\floatstyle{ruled}
\newfloat{listing}{tb}{lst}{}
\floatname{listing}{Listing}
\title{Build Your Own Robot Friend: An Open-Source Learning Module \\
for Accessible and Engaging AI Education}
\author{
    Zhonghao Shi\textsuperscript{\rm 1}\equalcontrib, 
    Allison O'Connell\textsuperscript{\rm 1}\equalcontrib, 
    Zongjian Li\textsuperscript{\rm 1}\equalcontrib, 
    Siqi Liu\textsuperscript{\rm 1}, 
    Jennifer Ayissi\textsuperscript{\rm 1},\\ 
    Guy Hoffman\textsuperscript{\rm 2}, 
    Mohammad Soleymani\textsuperscript{\rm 1},
    Maja Matarić\textsuperscript{\rm 1}
}
\affiliations{
    \textsuperscript{\rm 1}University of Southern California\\
    \textsuperscript{\rm 2}Cornell University\\

    \{zhonghas, amy.dell, liusiqi, ayissi, mataric\}@usc.edu\\
    \{hoffman\}@cornell.edu\\
    \{lizongjian, soleymani\}@ict.usc.edu
}

\usepackage{bibentry}

\begin{document}

\maketitle

\begin{abstract}

As artificial intelligence (AI) is playing an increasingly important role in our society and global economy, AI education and literacy have become necessary components in college and K-12 education to prepare students for an AI-powered society. However, current AI curricula have not yet been made accessible and engaging enough for students and schools from all socio-economic backgrounds with different educational goals. In this work, we developed an open-source learning module for college and high school students, which allows students to build their own robot companion from the ground up. This open platform can be used to provide hands-on experience and introductory knowledge about various aspects of AI, including robotics, machine learning (ML), software engineering, and mechanical engineering. Because of the social and personal nature of a socially assistive robot companion, this module also puts a special emphasis on human-centered AI, enabling students to develop a better understanding of human-AI interaction and AI ethics through hands-on learning activities. With open-source documentation, assembling manuals and affordable materials, students from different socio-economic backgrounds can personalize their learning experience based on their individual educational goals. To evaluate the student-perceived quality of our module, we conducted a usability testing workshop with 15 college students recruited from a minority-serving institution. Our results indicate that our AI module is effective, easy-to-follow, and engaging, and it increases student interest in studying AI/ML and robotics in the future. We hope that this work will contribute toward accessible and engaging AI education in human-AI interaction for college and high school students.
\begin{figure}[t!]
  \centering
  \includegraphics[scale=0.35]{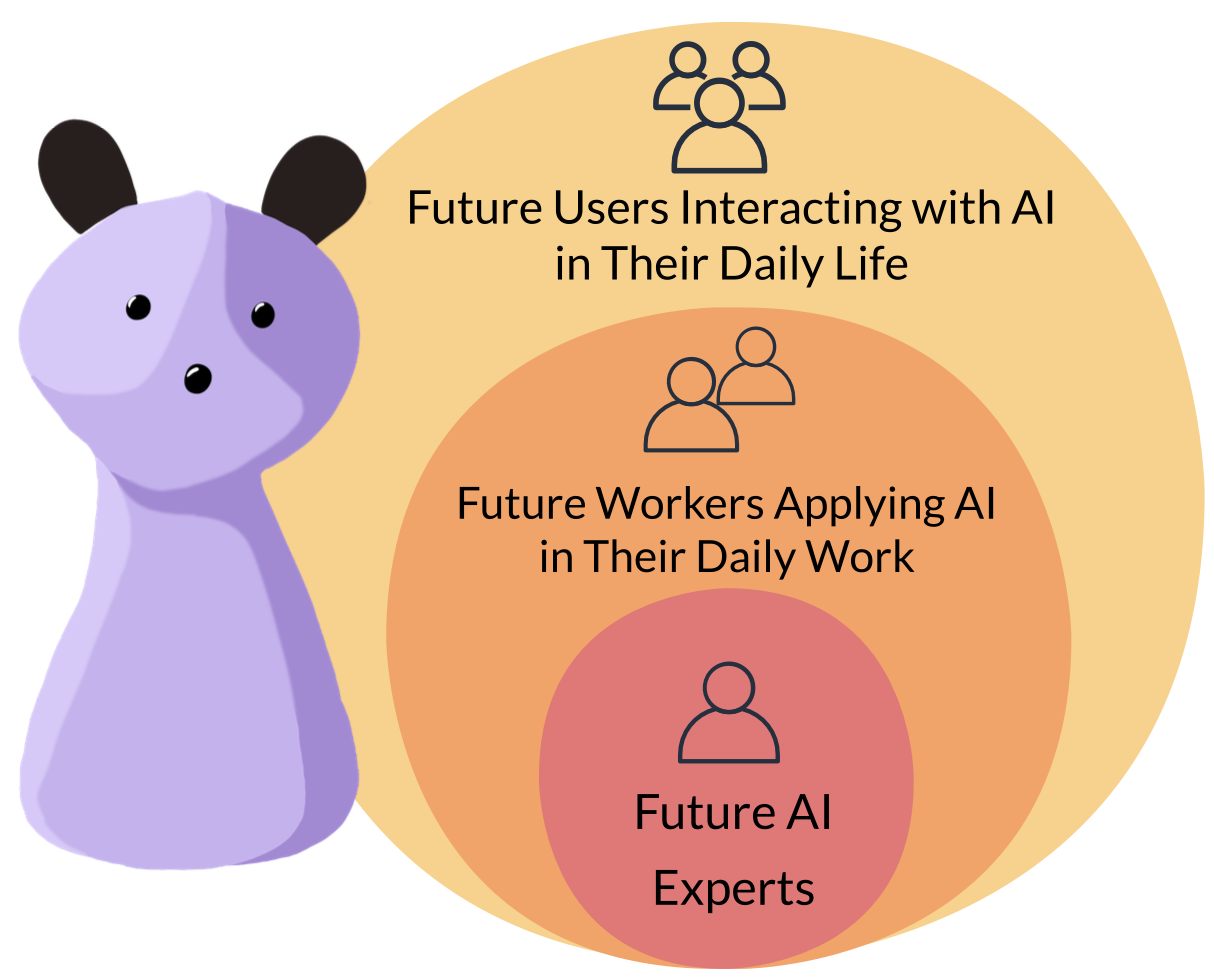}
  \caption{Our AI learning module aims to provide accessible and engaging AI education and literacy, not only for future AI researchers but also for everyone who may use or encounter AI in their daily life}
\label{fig:all}
\end{figure}

\end{abstract}

\section{Introduction}
\label{sec:intro}

From entertainment to healthcare and education, AI-powered technologies are increasingly integrated into our daily lives. 
Large-language models (LLM) such as ChatGPT~\cite{OpenAI2023GPT4TR} have shown tremendous potential in improving work efficiency in various industries, from software engineering~\cite{surameery2023use} to script editing and generation~\cite{hill2023chat}. Self-driving companies like Waymo and Cruise have deployed fleets of driverless taxis, which could change the way people commute every day. Despite these promising developments, as with similarly major technological advances throughout history, advances in AI may also further broaden the gap in inequality and leave behind users without sufficient AI literacy~\cite{dahlin2023social}. For example, users of LLM-based services may unintentionally leak out their proprietary or personal data if they do not have a basic understanding of how LLMs use and learn from their data~\cite{gupta2023chatgpt}. Without a sufficient understanding of AI ethics and fairness, users may blindly trust products such as AI-based resume screening services, resulting in unintended discrimination against marginalized groups~\cite{kelan2023algorithmic}. AI services such as Tesla's full self-driving feature may be misunderstood and used incorrectly, with potentially life-threatening consequences ~\cite{nordhoff2023mis}. For these reasons, we believe it is important to provide accessible and engaging AI education and literacy, not only for future AI researchers but also for everyone who may use or encounter AI in their daily life, as shown in Figure~\ref{fig:all}.

\begin{figure}[t!]
  \centering
  \includegraphics[scale=0.30]{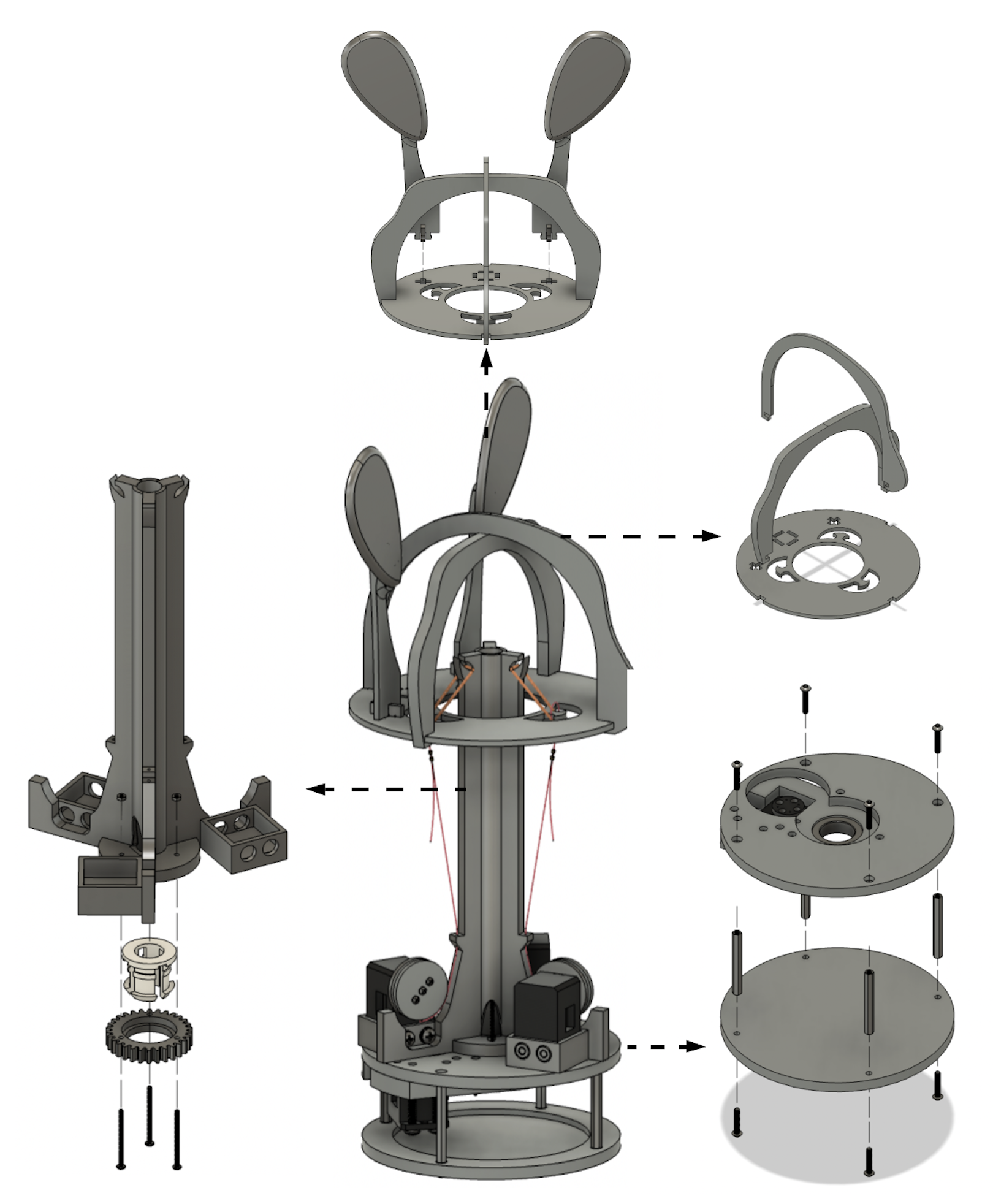}
  \caption{\textbf{Enabling Better Affordability:} Based on the robot platform proposed in \citet{suguitan2019blossom}, we simplified the robot's open-source hardware design for ease of assembly (reducing the number of fabricated components from 63 to 20) with only 3D-printed components. These improvements shorten the robot assembly time and make the robot more affordable and accessible.}
\label{fig:blossom}
\end{figure}


To provide more accessible AI education for all, the learning modules need to be affordable for students from all socio-economic backgrounds and customizable for learners with different learning goals. Many existing robot platforms have been validated as successful resources for AI/ML and robotics education. Platforms such as the NAO robot~\cite{shamsuddin2011humanoid} developed by Aldebaran or the Poppy robot~\cite{lapeyre2014poppy} have already been adapted to both K-12 and college classrooms to support better STEM learning~\cite{mubin2013review}. However, the cost of these robot platforms (NAO: over \$10k; Poppy robot: over \$7k) is not yet affordable for schools with limited resources. Alternatively, there are many commercial toy-like educational robot kits such as LEGO Boost (under \$300)~\cite{benedettelli2018lego} or Loona (under \$500)~\cite{northfield2022bizarre} available on the market. These kits are relatively inexpensive and are usually paired with a graphical programming user interface (UI). They have been shown to be a great, affordable resource for learning introductory AI topics. However, their one-size-fits-all closed-source design results in fixed learning goals, and hard-to-customize robot hardware and software. In this work, to enable accessibility for all learners, we aim to design an AI learning module that is affordable (under \$300) for students from all socio-economic backgrounds, provides open-source materials, documentation, and video tutorials, and is customizable and extendable to meet a wide range of learning goals.

Furthermore, existing AI learning modules focus mainly on AI agents without any social or expressive capabilities, so they have not yet considered human-centered AI and AI ethics. In robotics, the existing curricula and learning modules have already extensively covered conventional robotic tasks such as robot localization and manipulation~\cite{bers2014computational, correll2012one}. Similarly, in machine learning education, courses and learning modules mainly focus on the technical aspects of the machine learning pipeline, typically using datasets that are not human-centered~\cite{sanusi2023systematic}. However, as human-centered AI services become more prominent in daily work and life, it becomes important for AI education to cover topics on human-centered AI. In our proposed learning module, we aim to introduce students to human-centered AI topics such as human-robot interaction, human perception, affective computing, and AI ethics.

We propose an open-source, affordable, and customizable learning module for accessible AI/ML and robotics education for university and high school students. To evaluate our learning module, we conducted a two-day usability testing workshop with 15 college students recruited from a minority serving institution; our findings indicate that our learning module is effective, easy-to-follow, engaging, and could potentially encourage students to learn more about AI/ML and robotics in the future. The main contributions of this learning module are as follows:

- \textbf{Affordability:} Based on the robot platform proposed in \citet{suguitan2019blossom}, we significantly simplified the robot's open-source hardware design (reducing from 63 to 20 components) for ease-of-assembly with only 3D-printed components to improve accessibility for educational use. These improvements shorten the robot assembly time and make the robot more affordable and accessible.

- \textbf{Customizability:} Detailed documentations and video tutorials were created with the updated open-source software and hardware materials in the learning module, to make it easy for learners to not only create their own base version of the robot, but also to be creative and customize the robot with extended capabilities of their choice.

- \textbf{Human-Centered Focus:} We placed the AI learning module in the context of building one's own personal social robot and included educational content to cover topics related to human-centered AI, including human-robot interaction, human perception, affective computing, and AI ethics.

The rest of this paper is organized as follows. In~\nameref{sec:related_work}, we review prior work on learning platforms and modules for robotics and machine learning. We discuss the gaps in existing resources that our learning module aims to address. In~\nameref{sec:learning_module}, we present the details of our learning module and discuss how we aim to achieve more accessible and customizable AI education. In~\nameref{sec:evaluation_study}, we describe the design of our evaluation study to test the usability of our learning module. In~\nameref{sec:results}, we analyze the quantitative and qualitative data collected in our evaluation study. In~\nameref{sec:discussion}, we summarize the main takeaways from the study and present validated contributions and areas for improvements of our learning module. The~\nameref{sec:conclusion} wraps up the paper.

\section{Related Work}
\label{sec:related_work}

\subsection{Platforms for Robotics Education}

~\citet{mataric2007materials} define two general categories of educational robots: pre-built and do-it-yourself (DIY). Pre-built robots come fully assembled and ready to use in the classroom but are usually more expensive. DIY robots are less expensive, but require significant experience with hardware and electronics. Later reviews further separate educational robots into robotic kits, which allow students to create, build, and/or program robots, and social robot platforms, which interact with students toward a desired educational outcome (e.g., programmable robots, robot tutors) ~\cite{pachidis2019social}. Popular robot kits include LEGO Mindstorm ~\cite{legoMindstorms}, Boe-bot ~\cite{boebot}, and the VEX Robotics Ecosystem ~\cite{vexRobotics}. Socially expressive robots include the NAO~\cite{shamsuddin2011humanoid}, Keepon robots~\cite{belpaeme2018social}, QT robots~\cite{costa2017socially}, and the original Blossom robot~\cite{suguitan2019blossom}. 

While much research has explored the use of robot kits and socially expressive robots in the classroom, very few platforms fit into both categories. The Robotis BIOLOID series~\cite{han2008educational}, First Robotics~\cite{welch2007effect} and Botball robot~\cite{stein2002botball} are robot kits for robotics education and educational robotics competitions. The Poppy project designs open-source robots, including a humanoid robot, that can be assembled from 3D printed parts and off-the-shelf hardware ~\cite{lapeyre2014poppy}. The Buddy open-source robot is a 3D Printed Arduino-based social robot \cite{buddyRobot}. However, these and similar platforms are either costly, have limited social expressivity, or require technical skills that exceed what can be expected from K-12 students and teachers. 

The original Blossom~\cite{suguitan2019blossom} robot has the potential to be further modified and improved for education because of its inexpensive and expressive open-source design. In this work, we adapted Blossom for the classroom by updating the original 2D laser-cut design to 3D printed components and by significantly reducing the number of fabricated components for easier construction. Finally, we created new and more comprehensive documentation for the modified design to make our simplified Blossom robot an accessible platform for more accessible and engaging AI education.

\subsection{Accessible Machine Learning Education}

There has been an extensive body of work focusing on enabling more accessible and engaging machine learning education for college and K-12 students. \citet{gresse2021visual} presented visual tools to help students learn about the development pipeline of machine learning models; their study showed that these visual tools can effectively enhance student understanding of the learning materials. \citet{priya2022ml} designed a game to teach a general overview of three ML concepts (supervised learning, gradient descent and K-nearest neighbor (KNN) classification) in an incremental fashion. Their user study showed that the game can result in more effective and engaging learning of those ML concepts than conventional methods involving textbook learning for students in secondary school. Virtual reality has also shown promise as a supplementary tool to deliver learning materials to students in a more engaging fashion~\cite{jiang2021virtual}.

Despite the abundance of resources for AI/ML education, the majority of existing learning modules have not yet incorporated human-centered AI with social capabilities. \citet{alvarez2022socially} showed that existing AI/ML education tends to focus on the technical details of AI/ML. However, failing to connect AI/ML content to human-centered use cases could lead to lower participation from underrepresented students~\cite{ceci2014women}. In this work, by designing the learning module around a socially assistive robot, we aim to place an emphasis on human-centered and social components of AI to help students learn about topics such as human-robot interaction, human perception, affective computing, and AI ethics.

\section{Learning Module Design}
\label{sec:learning_module}
Our learning module consists of three parts. In Part 1, we focus on guiding each student in building their own Blossom robot (our modified and simplified version) from the ground up and learning how to program it in Python. During this process, the student also has the opportunity to customize Blossom's appearance and functionality based on their own interests. In Part 2, the student goes through a standard machine learning (ML) pipeline for human perception to learn how ML models are trained and evaluated in affective computing tasks. The tutorial will also cover AI ethics and algorithmic fairness. In Part 3, the student applies the ML models pre-trained for head pose tracking and gesture recognition to complete a basic human-robot interaction program, in which Blossom recognizes the user's head pose and imitates the user's head movements. Each part is described in more detail next.

\subsection{Part 1: Robotics}

\subsubsection{Build your own robot friend.}
Students begin the learning module by constructing their own robot. This activity was designed to introduce students to several robotics disciplines, including mechanical and electrical engineering. The robot's design is based on the Blossom open-source robotics research platform developed by the HRC2 Lab at Cornell University \cite{suguitan2019blossom}. We significantly modified Blossom's design to make the robot easier for novice users to assemble and more affordable for all users.

The original Blossom robot features a rigid inner structure covered by a soft exterior. Tensile mechanisms actuate the 5-DoF internal structure. The rigid components were designed to be laser-cut from wood sheets and glued together in layers to form the 3-dimensional components. The recent decline in the price of 3D printers has enabled many schools to own or have access to a 3D printer. For this reason, we updated the modified Blossom components to CAD models for 3D printing, shown in Figure~\ref{fig:blossom}. Updating the design from laser cutting to 3D printing also allowed us to take advantage of the additional dimension that 3D printing offers. No longer constrained to flat components, we consolidated several individual laser-cut pieces into a few unified components that can be printed without supports. In total, we decreased the number of printed components from 63 to 20, making the robot much less complex to construct.

We also created a new comprehensive guide to constructing the modified Blossom. The guide features 3D exploded diagrams of each step of construction with all parts labeled, giving a clear illustration of how the components should be assembled. We created equally detailed documentation for the robot's exterior, along with several opportunities for students to customize and extend Blossom, described in the following sections.

\subsubsection{Be creative with your robot friend's outfit!}

\begin{figure}[t!]
  \centering
  \includegraphics[scale=0.22]{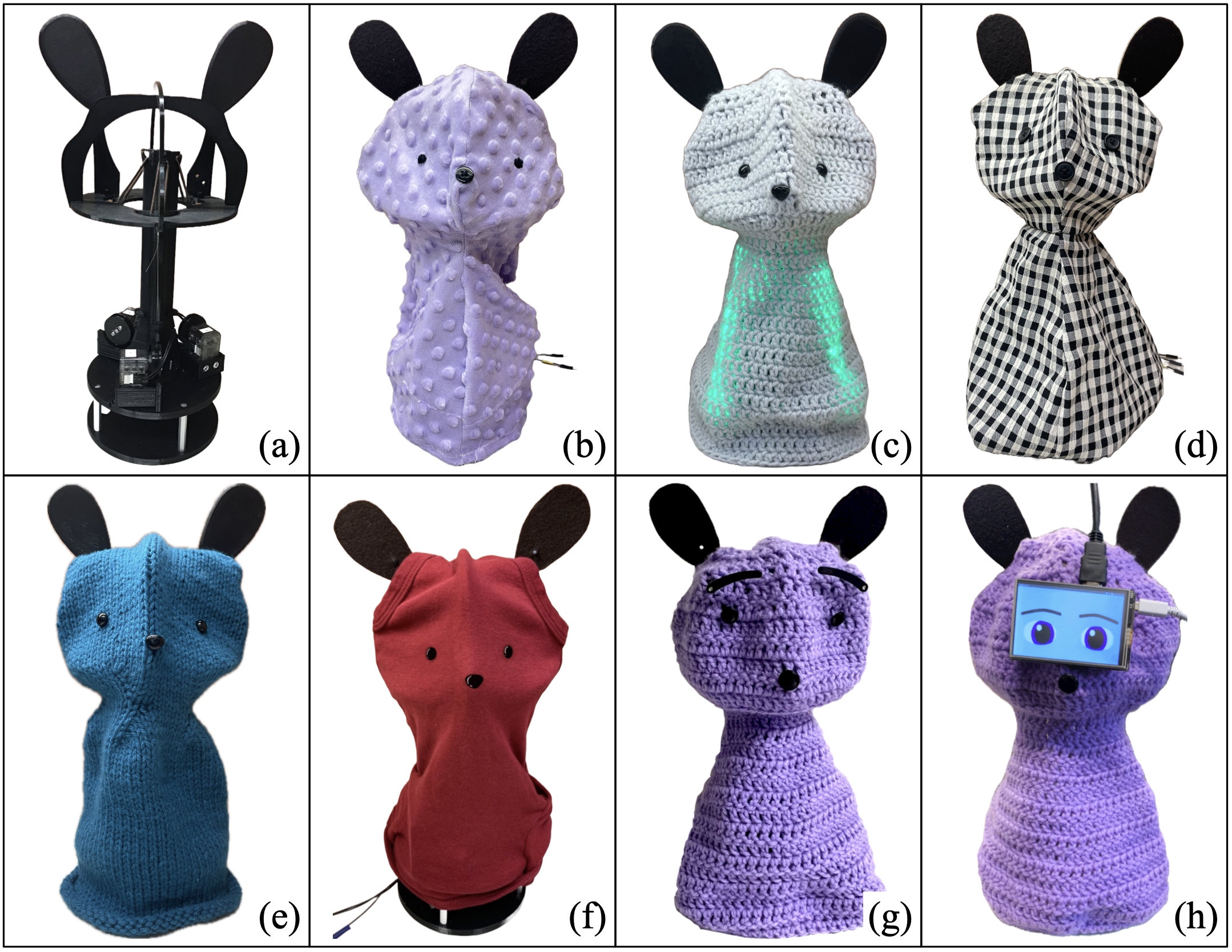}
  \caption{\textbf{Enabling Better Customizability:}
  Detailed documentation and video tutorials were created with the updated open-source software and hardware materials in the learning module, helping learners build their robot and be creative with its customizable design. A variety of exterior design tutorials and ideas to extend the platform gives students agency in their Blossom's design.}
\label{fig:collage}
\end{figure}

For the robot's exterior, we created several new cover designs for students with varied skill sets. The original Blossom platform features a knit or crocheted exterior that covers the rigid skeleton. To make the project more accessible and engaging, we designed more robot covers using a variety of materials and fabrication methods and coupled those designs with simple, user-friendly instructions. By encouraging personalization of the robot's exterior, we hope to provide more engaging and personalized learning experiences for each student.

To expand students' options for constructing the modified Blossom's exterior, we  designed five new covers using various materials and fabrication methods, as follows:

\begin{itemize}
    \item \textbf{Fabric Exterior From Repurposed Baby Onesie}
For users who want to avoid handcrafting a Blossom exterior, we designed a simple Blossom cover constructed from an off-the-shelf baby onesie, shown in Figure \ref{fig:collage}f. This design is inexpensive, can be quickly constructed, and requires no sewing, making it ideal for one-day workshops or for producing many Blossoms quickly. The user is still able to exercise creativity when adding facial features to Blossom.
    \item \textbf{Sewn Fabric Exterior}
We created two sewing patterns for Blossom's exterior, shown in Figure \ref{fig:collage}b and \ref{fig:collage}d. Each pattern requires less than one yard of fabric, cut into four or eight pieces that can be sewn together with a sewing machine or by hand. Suggestions for where facial features should be attached are marked on each pattern, but the user is free to explore other configurations for a customized appearance.
    \item \textbf{Knitted Yarn Exterior}
The original Blossom design featured a knitted cover, but no written instructions were available to replicate it, only an image of the flattened cover annotated with height and width dimensions. To make it easier for knitters to create a cover for their modified Blossom, we designed a simple knitted cover, shown in Figure \ref{fig:collage}e, and wrote a traditional knitting pattern for users to follow. The pattern uses primarily knit stitches and is suitable for users with little to no knitting experience.
    \item \textbf{Crochet Yarn Exterior}
The original Blossom design also featured a crocheted cover but only limited written instructions and an image annotated with some dimensions were provided. To help users crochet a cover for their Blossom, we designed a simple crocheted exterior, shown in Figure \ref{fig:collage}c and wrote a traditional crochet pattern for users to follow. The pattern uses three basic stitches (double crochet, slip stitch, DC2TOG) and is suitable for users with little to no crochet experience.

\end{itemize}

\begin{table*}[!htp]\centering
\caption{\textbf{Human-Centered ML learning component:} Our learning module places an emphasis on human-centered AI topics such as human-robot interaction, human perception, affective computing, and AI ethics and fairness.}\label{tab: }
\scriptsize
\begin{tabular}{|p{5in}|l|l|}\toprule
Learning Tasks &Learning Goals \\\midrule
Get started with downloading the selected subset of the CelebA dataset and visually view the datasets &\multirow{4}{*}{Data Pre-Processing and analysis} \\
Learn about the differences between supervised and unsupervised; classification and regression & \\
load in the dataset with pre-defined Keras DataLoader or create a customized DataLoader for your CelebA dataset & \\
Use the "Smiling" attribute as the target label for our classification problem, and analyze the data distribution for "Smiling" in the training, evaluation and testing set of the dataset & \\
\hline
Learn about the Bias-Variance Trade-off and No Free Lunch Theorem in machine learning &\multirow{3}{*}{Model Training} \\

Learn about the basics about feedforward neural network and convolutional neural network & \\

Create a basic convolutional neural network and the training loop & \\
\hline

Learn about how machine learning model is evaluated &\multirow{2}{*}{Model Evaluation} \\

Create the evaluation script to test the trained model on the testing data & \\
\hline
Learn about AI ethics and algorithmic fairness &\multirow{3}{*}{AI Ethics and Fairness} \\

Code a couple of popular evaluation metrics used for evaluating algorithmic fairness & \\
Add the algorithmic fairness metrics in the evaluation of your "Smiling" classifier, and determine how fair your classfier is & \\
\bottomrule
\end{tabular}
\end{table*}

To make the cover construction accessible for students, we wrote a comprehensive set of instructions for each of the five Blossom exterior designs, including high-resolution photographs depicting each step. The sewing, knitting, and crochet patterns are written in the standard format for each craft, making them easy for experienced craftspeople to follow. They also include glossaries and added instructions for beginners who may not be familiar with the techniques and terms. Using traditional fiber arts for fabrication encourages students to exercise creativity in crafting their robot cover and gives them greater agency in their Blossom's design

\subsubsection{Maybe give your friend a superpower?}

Beyond instructions to build the basic version of Blossom, the learning module encourages students to customize Blossom with an array of added functionalities. For example, the original Blossom design features a fifth motor mounted on the head platform that can add a degree of freedom to the robot. We removed this motor from the base version as it is hard to install, but we include examples of how students may use this motor to add expressivity and character to the robot. The added actuator can move a modified pair of ears on the robot's head or eyebrows on the robot's face (Figure \ref{fig:collage}g). A screen can also be mounted on the robot's head to give the robot an animated face (Figure \ref{fig:collage}h). We also show a version of the robot with RGB lights inserted underneath the crocheted exterior to create dynamic light effects (Figure \ref{fig:collage}c). 

\subsection{Part 2: Human-Centered AI, AI ethics and fairness}

In the learning module, we aim to cover the basics of the machine learning pipeline, including the following topics: 1) dataset pre-processing and analysis, 2) model training and evaluation, 3) AI ethics and fairness. By finishing this module, students will gain an introductory understanding of the human perceptual ML models used for computer vision and will gain more confidence in applying and programming with ML models to finish the rest of the learning module.

\subsubsection{Train your robot friend to see you.}

After students finish building and programming Blossom with basic movements, we next teach the students how their Blossom and other AI systems perceive human behaviors using computer vision ML models. Based on the official PyTorch tutorial in Colab~\cite{paszke2019pytorch}, instead of using the non-human-centered CIFAR10 dataset~\cite{krizhevsky2009learning}, we updated the tutorial so that students will train a convolutional neural network (CNN) on the human-centered CelebA dataset ~\cite{liu2015faceattributes} for a binary "Smiling" or "Not Smiling" classification task. This tutorial helps students gain introductory knowledge about the AI/ML pipeline and how their Blossom robot can see them.

\subsubsection{Help your robot friend learn about ethics and fairness.}

Having built a relatable robot, the tutorial next introduces students to some of the ethical issues related to AI. We adapted the recommendations of ethical AI published by the United Nations Educational, Scientific, and Cultural Organization (UNESCO) to discuss the 10 core principles for an ethical future of human-centered AI~\cite{unesco2021recommendation}. Specifically, we cover what algorithmic fairness is and the popular fairness metrics that have been actively used in the AI fairness literature, including: 

\vspace{10pt}
Demographic Parity:

\begin{equation}
P(\hat{Y} \mid S=0) = P(\hat{Y} \mid S=1)
\end{equation}

\vspace{5pt}
Equalized Odds: 

\begin{equation}
P(\hat{Y} \mid S=0, Y=y) = P(\hat{Y} \mid S=1, Y=y), y\in\{0, 1\}
\end{equation}

\vspace{5pt}
Equalized Opportunity:

\begin{equation}
P(\hat{Y} \mid S=0, Y=1) = P(\hat{Y} \mid S=1, Y=1)
\end{equation}

\vspace{10pt}

We define the notation of the dataset as $\{ (\mathbf{x}_i, s_i, y_i )_{i=1}^{N} \}$, and the number of total data instances is $N$. For a given instance $i$, the non-sensitive features of the instance are denoted as $\mathbf{x}_i \in \mathbb{R}^d$. The label of the classification task is denoted as $s_i \in \{0,1\}$. In our tutorial, our labels of the classification task are "Smiling" and "Not Smiling." The sensitive attribute is denoted as $s_i \in \{0,1\}$, and gender is used as our sensitive attribute in our tutorial. $X$, $Y$, $S$ and $\hat{Y}$ are used to represent the corresponding random variables of ${x}_i$, ${y}_i$, ${s}_i$, and $\hat{y}$.

At the end of the tutorial, the students are guided to code a function for each of the three fairness metrics and include them in the model evaluation for their binary "Smiling" classifier, to showcase how unfair the models are without fairness interventions. This tutorial aims to highlight the underlying bias of ML models and the importance of AI fairness for more trustworthy human-centered AI solutions.

\subsection{Part 3: Software Engineering with AI/ML and Robotic Libraries}

\subsubsection{Teach your friend how to see you.}

After students have built their robot and learned about ML models, the final goal of this learning module is to complete a basic human-robot interaction program in which Blossom recognizes the user’s head pose and imitates the user’s head movements. To enable the robot to perceive the human, we ask the students to integrate the robot control library with another open-source library called OpenSense, an open source research platform for real-time multimodal acquisition and recognition of social signals released by~\citet{stefanov2020opensense}.

\begin{figure}[t!]
  \centering
  \includegraphics[scale=0.24]{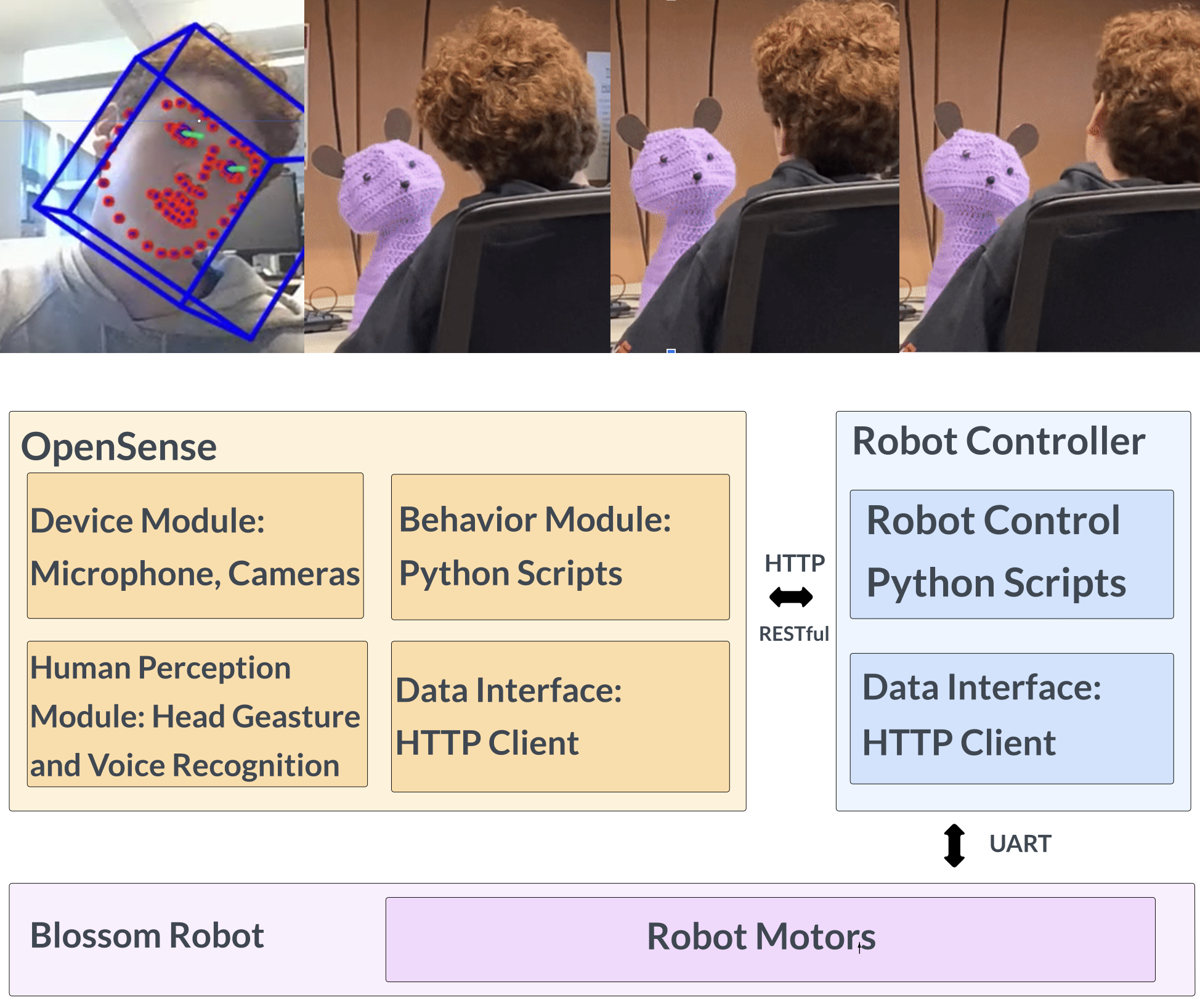}
  \caption{System architecture of the final program that students develop in this learning module, where the robot is programmed to recognize the user’s head pose and imitate the user’s head movements.}
\label{fig:program}
\end{figure}

To program OpenSense, we offer a graphical user interface (GUI) editor. Using this editor, students can pick and manage modules for data capture, perception, understanding, behavior management, and data preservation and communication. We developed written documentation and video tutorials for OpenSense and the integration between OpenSense and the Blossom robot for students to follow as they finish the human-robot interaction program in this learning module. For those who are interested in learning more about the source code and customizing the functionalities, the editor also allows for the quick insertion of customized rules and logic in Python.

\begin{figure}[t!]
  \centering
  \includegraphics[scale=0.55]{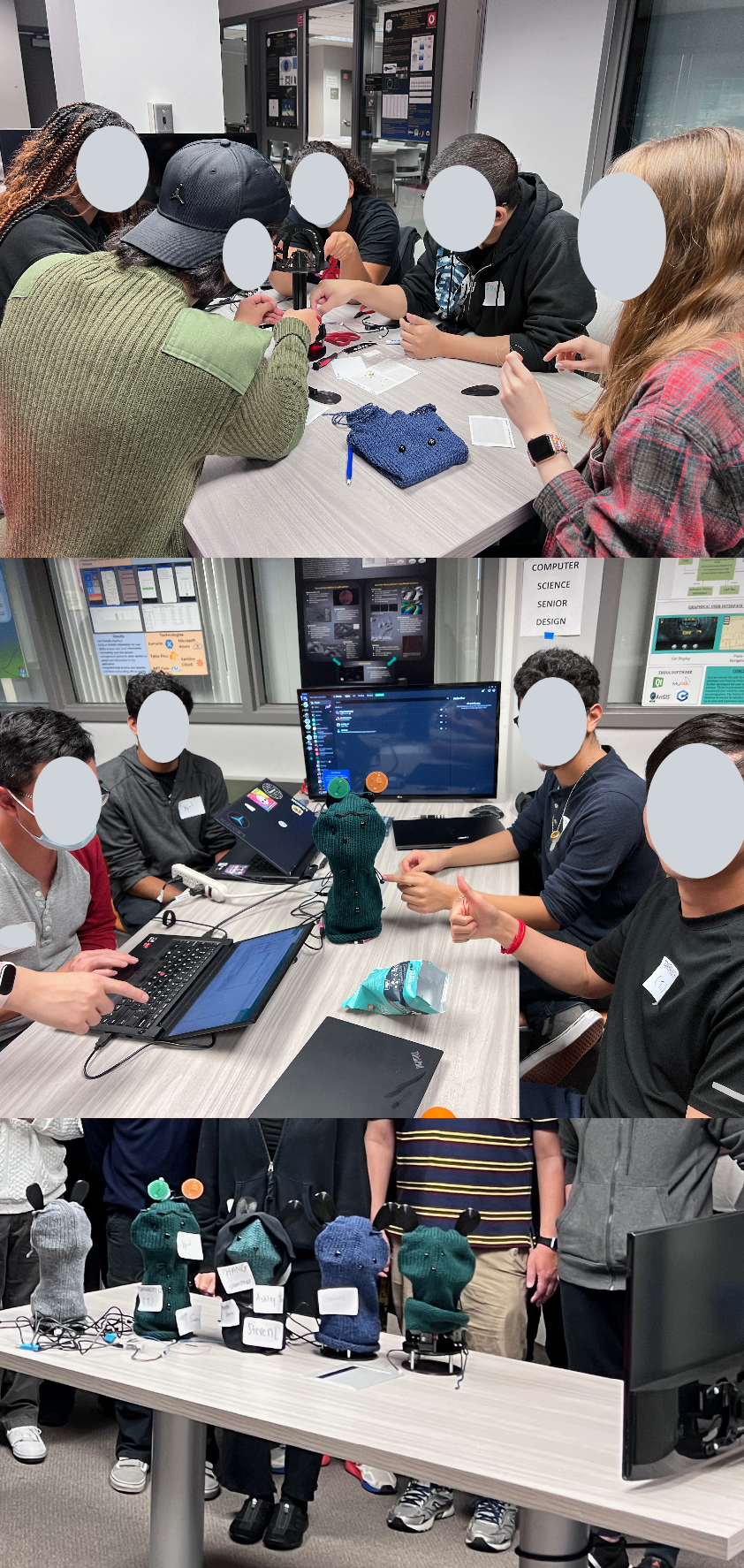}
  \caption{\textbf{Human-centered robotic learning component:} By designing the learning module around a socially assistive robot, we aim to engage the students in learning about topics on human-centered and social components of AI.}
\label{fig:workshop}
\end{figure}

\subsubsection{Teach your robot to imitate you.}


Once the program is set up, students should have correctly connected the modules for data capture, human perceptual modules, behavior modules, and the data interface. After that, students can run the program to enable the robot to follow their head movements. When the user nods, the robot nods in return; when the user shakes their head, the robot does the same. All our software implementations are open-source and easy to use and modify, so they are also customizable and extensible to fit different students with their individual learning goals.

\section{Evaluation Study Setup}
\label{sec:evaluation_study}

We conducted a two-day in-person participatory design workshop to gather feedback and evaluate our learning module for teaching AI/ML and robotics. Due to the time constraints of the workshop, we only evaluated Parts 1 and 3 of our learning module during our workshop. We conducted the workshop at an Asian American and Native American/Pacific Islander Serving Institution (AANAPISI) and Hispanic-Serving Institution (HSI). The workshop covered the hardware and software components of the Blossom robot (our modified version) and guided the participants through building and programming the robot to recognize, track, and imitate their head pose. 15 participants (10 male, 2 female, 1 non-binary, 2 did not disclose) participated in the workshop; 7 participants self-identified as Asian American or Native American/Pacific Islander, 4 participants self-identified as Hispanic, 1 participant self-identified as African American, and 3 did not disclose. The participants were randomly divided into four teams, and all teams successfully completed the workshop goals.

\begin{figure}[t!]
  \centering
  \includegraphics[scale=0.18]{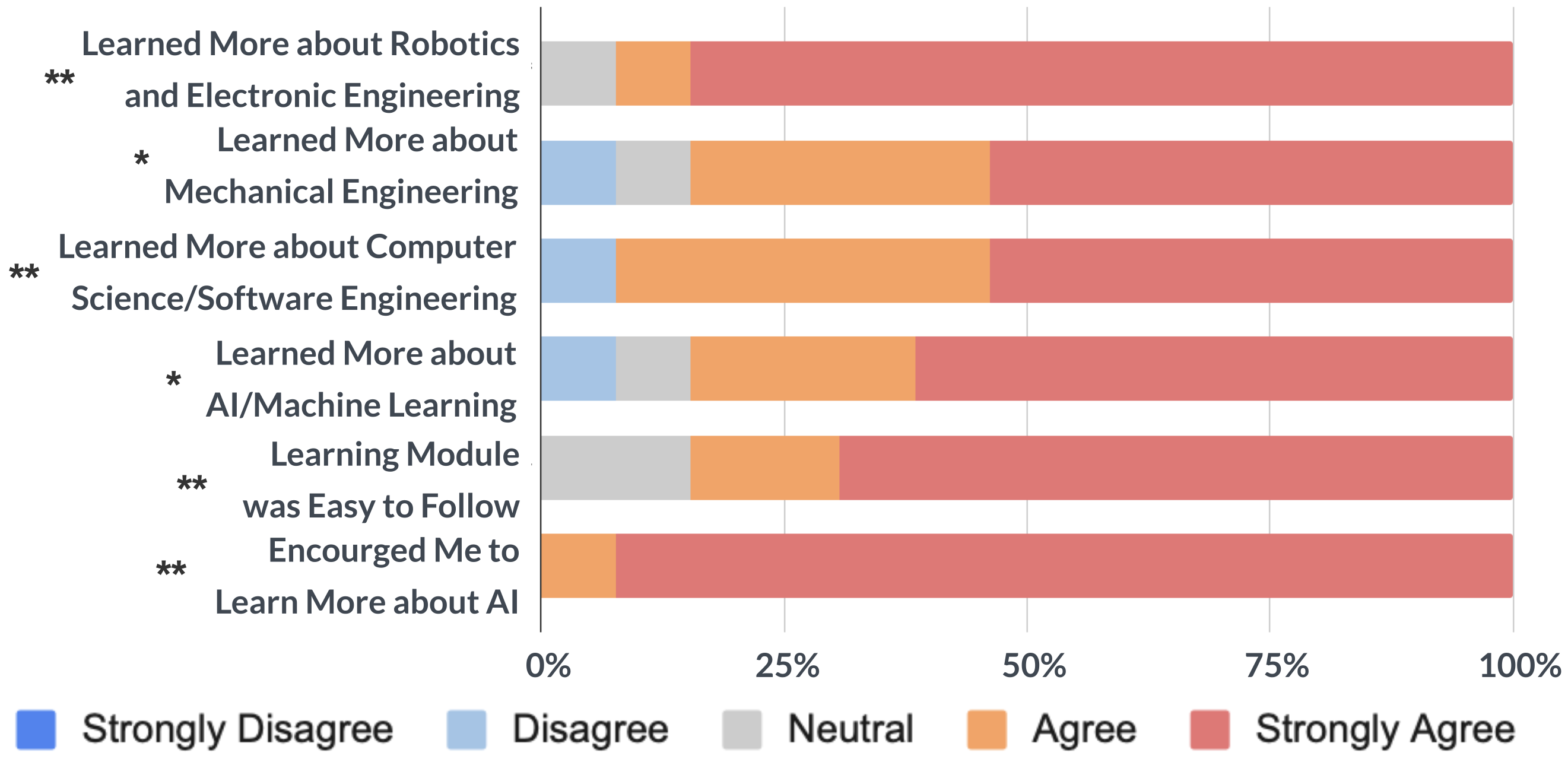}
  \caption{Our quantitative results indicate that our learning module is effective, easy-to-follow, engaging, and increases the students’ interests in further studying robotics and AI in the future (* = $p$ \textless .05, ** = $p$ \textless  .001). }
\label{fig:likert}
\end{figure}

At the end of the two-day workshop, we conducted a post-workshop survey to study how the participants perceived the usability of our learning module for teaching AI/ML and robotics. Of the 15 participants, 13 filled out the post-workshop survey. The survey included 5-point Likert scale questions to evaluate the student-perceived usability of our learning module. In addition, we also collected qualitative feedback for future improvements. In our quantitative analysis, a one-sample Wilcoxon signed rank test was used to determine whether the median of our collected responses was greater (more favorable) than or equal to 3 (neutral) on the 5-point Likert scale.

\section{Results}
\label{sec:results}

Based on the quantitative results, we found that 92\% of the participants believed that the workshop helped them learn more about robotics/electronic engineering (median = Strongly Agree (2), $p$ = .0004), computer science/software engineering (median = Strongly Agree (2), $p$ = .0008), AI/ML (median = Strongly Agree (2), $p$ = .0014), and mechanical engineering (median = Strongly Agree (2), $p$ = .0018). In addition, 85\% of the participants said that the process of our learning module was easy-to-follow (median = Strongly Agree (2), $p$ = .0009). 100\% of the participants believed that the workshop helped to encourage them to study more about robotics and AI in the future (median = Strongly Agree (2), $p$ = .0001).

\section{Discussion}
\label{sec:discussion}

As shown in Figure~\ref{fig:likert}, our quantitative results indicated that our learning module with our modified version of the Blossom robot could provide effective, easy-to-follow, engaging AI learning that increases the students’ interests in further studying robotics and AI in the future. At the end of the two-day workshop, all four teams successfully finished the learning module by completing the human-robot interaction program. Despite coming from different academic backgrounds, the students demonstrated no difficulty in following the tutorials and finishing the required tasks. Similarly to the quantitative results, the qualitative results further validated that the students were able to gain introductory knowledge about AI and domains that are outside their own majors. While preparing for the workshop, we also verified that the robot could cost under \$250 if the 3D-printed parts were printed at scale, and this validated the affordability and accessibility of our module.

By designing the learning module around a social robot, the students found the learning content more relatable and engaging. At the beginning of the learning module, when we first introduced the students to human-centered AI and robotics, we compared the Blossom robot with Baymax~\cite{murphy2022safe}, a fictional robot character from the popular movie "Big Hero 6". One of the students wrote in the survey: "I liked the reference to Baymax and am excited to learn more about the applications of these types of robots." Another student wrote: "This is great for those who just started learning AI and are interested in new project or get some more motivation." 

\section{Conclusion and Future Work}
\label{sec:conclusion}

In this work, we developed and evaluated a learning module to provide more accessible and engaging AI/ML and robotics education in college and high school education. We significantly simplified the original Blossom robot design to make it easier to build and made it more affordable and accessible for educational use. We also validated that the cost for each robot can be under \$250 if the components are printed as scale, or in-house. In addition, we included new and detailed documentation, video tutorials, and open-source hardware/software materials, so the learning module is readily customized for different students and learning goals. Our two-day usability testing workshop showed that our module is effective and easy to follow and encourages students to explore more independent studies of AI and robotics. Finally, our human-centered focus with a social robot not only covered content that is currently rarely covered in AI learning modules but also enabled more engaging learning experiences. In continued work, we plan to conduct an evaluation workshop for Part 2 of our learning module and to evaluate the module with high school students to further validate and improve the module for students at different educational levels.

\section{Acknowledgements}
This research was supported by the National Science Foundation CISE Community Research Infrastructure Grant NSF-2233191.

\bibliography{aaai24}
\end{document}